\documentclass[%
 aip,
 amsmath,amssymb,
 reprint,%
]{revtex4-1}

\usepackage{graphicx}
\usepackage{dcolumn}
\usepackage{bm}

\usepackage[utf8]{inputenc}
\usepackage[T1]{fontenc}
\usepackage{mathptmx}
\usepackage{xcolor}
\DeclareUnicodeCharacter{2061}{}

\begin{document}

\preprint{AIP/123-QED}

\title[]{Extended Source of Indistinguishable Polarization-entangled Photons over Wide Angles of Emission}


\author{Salem F. Hegazy}
\email{shegazy@zewailcity.edu.eg.}
\affiliation{National Institute of Laser Enhanced Sciences, Cairo University, Giza 12613, Egypt}
\affiliation{Centre for Photonics and Smart Materials, Zewail City of Science and Technology, Giza 12578, Egypt}

\author{Salah S. A. Obayya}
\email{sobayya@zewailcity.edu.eg.}
\affiliation{Centre for Photonics and Smart Materials, Zewail City of Science and Technology, Giza 12578, Egypt}


\begin{abstract}
The generation of high-fidelity polarization-entangled photon pairs, to date, has been demonstrated on specific spatial modes or over relatively narrow apertures.
 We put forward and demonstrate an experimental scheme to extend the temporal and spatial indistinguishability of polarization-entangled photons over wide emission angles, which can be applied to cover the whole SPDC cone.
 Over such wide angular extent, while the time-delay map is almost flat which renders the conventional compensation via a birefringent element an appropriate approach, the relative-phase map –verified as a quadratic function– necessitates a tunable compensation paradigm.
Here, to do so, we employ a phase-only two-dimensional spatial light modulator (2D SLM) loaded by the complementary of the relative phase map to equalize the phase variations for one third of the noncollinear spontaneous parametric down conversion (SPDC) emission. 
After eliminating the temporal and spatial distinguishability over the 2D SLM area, a $97\%$ polarization visibility is verified for the entangled photon pairs scattered widely across the SPDC cone.
 %
\end{abstract}

\maketitle

In the spontaneous parametric downconversion (SPDC) process, portion of pump photons split into photon pairs created with energies strictly satisfying the energy conservation condition and scattered in directions promoted by the momentum conservation (or phase-matching) condition.
The SPDC photons are thus entangled by nature in energy-time \cite{hong1987measurement,franson1989bell} as well as in momentum-position \cite{mair2001entanglement,lima2006propagation,barbieri2007complete, walborn2010spatial,leach2010quantum}. However for the polarization, the remaining degree of freedom (DoF), its entanglement appears always as a consequence of subtle designs using either a single \cite{kwiat1995new,Hugues2006non,kim2006phase,ueno2012entangled} or more nonlinear crystals \cite{kwiat1999ultrabright,kim2001interferometric,ljunggren2006theory,hegazy2017orthogonal,Villar2018Experimental}.  

In either case, the wavefunctions of the two polarization possibilities (HH and VV for $|\phi^\pm\rangle$, or HV and VH for $|\psi^\pm\rangle$) can be made almost identical in magnitude, while a comprehensive picture about the indistinguishability in the space and time domains is provided by the directional-spectral relative-phase function. Generally, a flatter relative-phase function in some domain implies higher indistinguishability in its conjugate Fourier-transform domain
 \cite{hegazy2017orthogonal,Hegazy2017CLEO}. 
Therefore, the spatial and temporal compensations actually correct (or flatten) the varying relative-phase function in direction and frequency. Several works previously addressed spatial and temporal compensations at specific spatial modes \cite{kwiat1995new,nambu2002generation,ljunggren2006theory,trojek2008collinear} or over relatively narrow aperture \cite{altepeter2005phase,rangarajan2009optimizing,cialdi2010programmable} by the use of additional birefringent elements. 
Over such limited spatial extents, well-designed birefringent elements can effectively clean for the linear dependence of the relative phase function and restore the indistinguishability \cite{hegazy2015tunable}. 
%
This is not the case when photon pairs are collected over relatively wide angular extents, where higher-order contributions of the relative-phase function drastically
reduce the effectiveness of linear methods.

In this paper, we present and demonstrate experimentally a method for eliminating the spatial and temporal distinguishability of a polarization-entangled state over wide emission angles.
We initially derive the two-photon state emitted by a noncollinear SPDC; including the directional-spectral relative-phase function all over the SPDC cone.
It is noticed that while the time-delay map (the spatial distribution of spectral relative-phase gradient) has a negligible change over wide angular extent, the directional relative-phase exhibits obvious quadratic profile. To restore the temporal and spatial indistinguishability, the compensation of the former can be effectively done in the conventional way; using a birefringent element. The latter is manipulated using a two-dimensional spatial light modulator (2D SLM) loaded with the inverted relative-phase profile. A fast yet accurate experimental technique is introduced to directly measure the relative-phase profile; dispensing with the quantum state tomography.
Consider a diagonally polarized pump beam illuminating a couple of crossed type-I crystals which are identically cut to promote production of degenerate photons in a noncollinear SPDC geometry. 
The pump beam is classically treated as a superposition of frequency and plane-wave  modes:
$E_{{p}}(\mathbf{x},z,t)= \int d\omega _{{p}}d\mathbf{q}_{{p}}~A_{p}(\omega_{{p}};\mathbf{q} _{{p}})\exp i\left(\kappa _{{p}}z+\mathbf{q}_{{p}}. \mathbf{x}-\omega _{{p}}t\right) +\mathrm{c.c.}$,
with angular frequency $\omega _{p}$ and wavevector $\mathbf{k}_{p}=(%
\mathbf{q}_{{p}},\kappa _{{p}})$ where $\mathbf{q}_{{p}%
} $ are the transverse components along the coordinates $\mathbf{x}=(x,y)$, and $\kappa _{p}$ is the longitudinal component. 
The pump polarization components in horizontal/vertical (H/V) basis have a relative phase $\phi_p$ that can be manipulated by tilting a birefringent element; e.g., a quarter wave plate (QWP). 

Similarly, the produced SPDC two-photon emission can be expressed –based on the SPDC creation operators– as a joint spectral and spatial expansion in monochromatic planar waves with the angular frequencies $\omega_{1,2}$ and the transverse wave vectors  $\mathbf{q}_{\mathrm{1,2}}$ (subscripts 1, 2 denote signal and idler photons, respectively).
The longitudinal wavevectors of the interacting waves can be expressed as dictated by Maxwell's equations for ordinary (\textit{o}) and extraordinary (\textit{e}) polarization as
\cite{born2013principles,walborn2010spatial,hegazy2017orthogonal}
\begin{equation}
\begin{split}
&\kappa _{{j}}^{o }(\omega_{{j}};\mathbf{q}_{{j}})=\sqrt{( {\omega_{{j}} ~n_{{j}}^{o}}/{c}) ^{2}-\vert \mathbf{q}_{{j}}\vert
^{2}}\\
&\kappa _j ^{e(H,V)}(\omega_j;\mathbf{q}_j)
=   q_{j(x,y)} \tan\rho^\perp_j \\
&~~~~~~~~~~~~+ n_j^{e^\perp} \left[ \left(\tfrac{\omega _j}{c} \right) ^{2} -         \tfrac{1}{(n_j^e)^2}   q_{j(y,x)}^2-       (\tfrac{{n_j^{e^\perp}}}{n_j^e n_j^o} )^2         q_{j(x,y)}^2 \right]^{\frac{1}{2}},\\
\end{split}
\label{eq:Longwavevector}
\end{equation}
where the subscript $j=p, 1, 2$, the superscripts $(H)$ and $(V)$ label the SPDC crystals with optic axes lying in the $xz$ (horizontal) and $yz$ (vertical) plane respectively, $c$ is the speed of light in space, $n_{{j}}^{o}$ and $n_{{j}}^{e}$ are principal values of refractive index, and $n_j^{e^\perp}$ and $\rho^\perp_j$ are  the refractive indices and walk-off angles of the extraordinary-polarized ray propagating along $z$ axis (perpendicular to the crystals interfaces).
The SPDC in the two crystals creates a two-photon state entangled in every degree of freedom; frequency, momentum, and polarization, which is expressed at small emission angles  \cite{migdall1997polarization} by the superposition
\begin{equation}
\begin{split}
\left\vert \psi \right\rangle   \sim  \mathbf{\int}d\omega _{1}d\omega _{2}d\mathbf{q}_{1}d%
\mathbf{q}_{2} &{\large \{}\Phi _\mathrm{HH}\left(\omega _{1},\omega _{2};\mathbf{q}%
_{1},\mathbf{q}_{2}\right)\left\vert H_1H_2
\right\rangle \\
&+~\Phi _\mathrm{VV}\left( \omega _{1},\omega
_{2};\mathbf{q}_{1},\mathbf{q}_{2}\right) \left\vert V_1V_2
\right\rangle{\large \}}, 
\end{split}
\label{ent1}
\end{equation}
where $\Phi _\mathrm{HH,VV}\left(\omega _{1},\omega _{2};\mathbf{q}_{1},\mathbf{q}_{2}\right)$ are the biphoton wavefunctions corresponding to the HH and VV possibilities.
 Assuming that the crystals are of infinite transverse extent and pumped by a non-depleted beam with the reflected waves at all crystal interfaces being negligible, the two-photon wavefunctions can be written as 
\begin{equation*}
\begin{split}
\Phi_\mathrm{HH} = ~&\tfrac{\chi^{(2)}}{\sqrt[]{2}} A_p(\omega_1+\omega_2;\mathbf{q}_1+\mathbf{q}_2) \int_{-2L}^{-L} dz~ e^{i \int_0^z dz^{\prime } \Delta \kappa(z^{\prime })}, \\
\propto ~& 
e^{-i L\left(\tfrac{1}{2}{\Delta \kappa_{V}^{ooe}} + \Delta \kappa_H^{eeo}  \right)} 
L
A_p(\omega_1+\omega_2;\mathbf{q}_1+\mathbf{q}_2) ~ \mathrm{sinc} \left( \tfrac{ \Delta \kappa_{V}^{ooe} L}{2}\right), \\
\Phi_\mathrm{VV} = ~&  e^{-i\phi_p}~ \tfrac{\chi^{(2)}}{\sqrt[]{2}} 
A_p(\omega_1+\omega_2;\mathbf{q}_1+\mathbf{q}_2) \int_{-L}^{0} dz~ e^{i \int_0^z dz^{\prime } \Delta \kappa(z^{\prime })}, \\
\propto ~&
 e^{-i\left(\phi_p+\tfrac{1}{2}{\Delta \kappa_{H}^{ooe}L} \right)}  
L
A_p(\omega_1+\omega_2;\mathbf{q}_1+\mathbf{q}_2)  ~\mathrm{sinc} \left(\tfrac{\Delta \kappa_{H}^{ooe} L}{2}\right), \\ 
\end{split}
\label{Eq: amplitudes}
\end{equation*}
where 
$L$ is the crystal length, $\chi ^{(2)}$ is the bilinear susceptibility, 
$~\Delta \kappa^{ooe}_{H,V} =\kappa _{p}^{e(H,V)}-\kappa _{1}^{o}-\kappa _{2}^{o}$ are  wavevector mismatches  within the interacting $(H)$ and $(V)$ crystals, 
$\Delta \kappa_H^{eeo}=\kappa _{p}^{o}-\kappa _{1}^{e(H)}-\kappa_{2}^{e(H)}$ is the mismatch of waves interacting in the first crystal when passing through the second one ($\Delta \kappa_H^{eeo} \gg \Delta \kappa^{ooe}_{H,V}$, therefore no downconversion is considered in the later case). 

The relative phase of the produced state $|\psi\rangle$ is then
\begin{equation}
\begin{split}
\vartheta 
=\mathsf{arg}\left\{  \Phi_\mathrm{VV} / \Phi_\mathrm{HH} \right\}=
\left[\tfrac{1}{2}({\Delta \kappa_{V}^{ooe} }-\Delta \kappa_{H}^{ooe}) + \Delta \kappa_H^{eeo}  \right] L -\phi_p,
\end{split}
\label{eq:relative_phase}
\end{equation}
which includes phase-matching terms and an initial phase term. We assume that the spatiotemporal amplitude of the pump beam is factorizable:
$A_p(\omega_p;\mathbf{q}_p)= ({\sqrt[]{\tau_p}}/{\sqrt[4]{\pi}})\,\exp [-\tfrac{1}{2} \tau_p^2 (\omega_{{p}}-\omega_p^0)^2 ] \,\,\, \delta(\mathbf{q}_p),$ which expresses a polychromatic plane wave with a coherence time $\tau_p$. Therefore, the biphoton spatial characteristics can be fully determined with reference to the signal photon alone ($\mathbf{q}_2 = -\mathbf{q}_1$). This assumption implies also that $\Delta \kappa_{V}^{ooe} \approx\Delta \kappa_{H}^{ooe}\equiv\Delta \kappa^{ooe}$ as dictated by Eq. (\ref{eq:Longwavevector}); and consequently $|\Phi_\mathrm{HH}|\approx|\Phi_\mathrm{VV}|$. By substituting Eq. (\ref{eq:Longwavevector}) into Eq. (\ref{eq:relative_phase}) and ignoring the insignificant higher-order terms, the relative phase writes
\begin{equation}
\begin{split}
    \vartheta  (\omega _{1},\omega _{2};  \mathbf{q}_{1},  & -\mathbf{q}_{1}) \approx\\
\Big \{ & \frac{(n_p^o - n_1^{e^\perp}) \omega_1 + (n_p^o -  n_2^{e^\perp}) \omega_2}{c}\\
&- q_{1,x} (\tan⁡ \rho_1^\perp   -   \tan⁡ \rho_2^\perp ) \\
&+  \frac{c q_{1,x}^2}{2}    \Big[ \frac{(n_1^{e^\perp})^3}{\omega_1  (n_1^e )^2 (n_1^o)^2}  + \frac{(n_2^{e^\perp} )^3}{\omega_2  (n_2^e )^2 (n_2^o )^2}\Big] \\
&+ \frac{c q_{1,y}^2}{2} \Big[ \frac{n_1^{e^\perp}}{\omega_1 (n_1^e )^2} + \frac{n_2^{e^\perp}}{\omega_2 (n_2^e )^2 }\Big] \Big\} L - \phi_p,
\end{split}
\label{eq:relative_phase_approx}
\end{equation}
which is a quadratic function in the transverse wavevector components of the SPDC photons.
 Moreover, the first-order derivatives of the relative phase $\vartheta$ with respect to $\omega_1$ and $\omega_2$ determine the time delay \textit{map} between HH and VV possibilities for the signal and idler photons, respectively [that is, for a biphoton emission in the directions specified by $(\mathbf{q}_{1},-\mathbf{q}_{1})$]. Equation (\ref{eq:relative_phase_approx}) thus presents closed forms for the relative-phase and time-delay maps which give theoretical predictions equivalent to those of the iterative approach given in Ref. \cite{hegazy2017relative}. For convenience, the components of $\mathbf{q}_1$ are substituted by the free-space emission angles in $xz$ (horizontal) and $yz$ (vertical) planes as $q_{1x,y}= \tan \theta_{x1,y1}~\sqrt[]{(\omega_1/c)^2-|\mathbf{q}_1|^2}$ (similarly, $\mathbf{q}_2$). 

\begin{figure*} [t]
   \begin{center}
   \begin{tabular}{c} 
   \includegraphics[width=12cm]{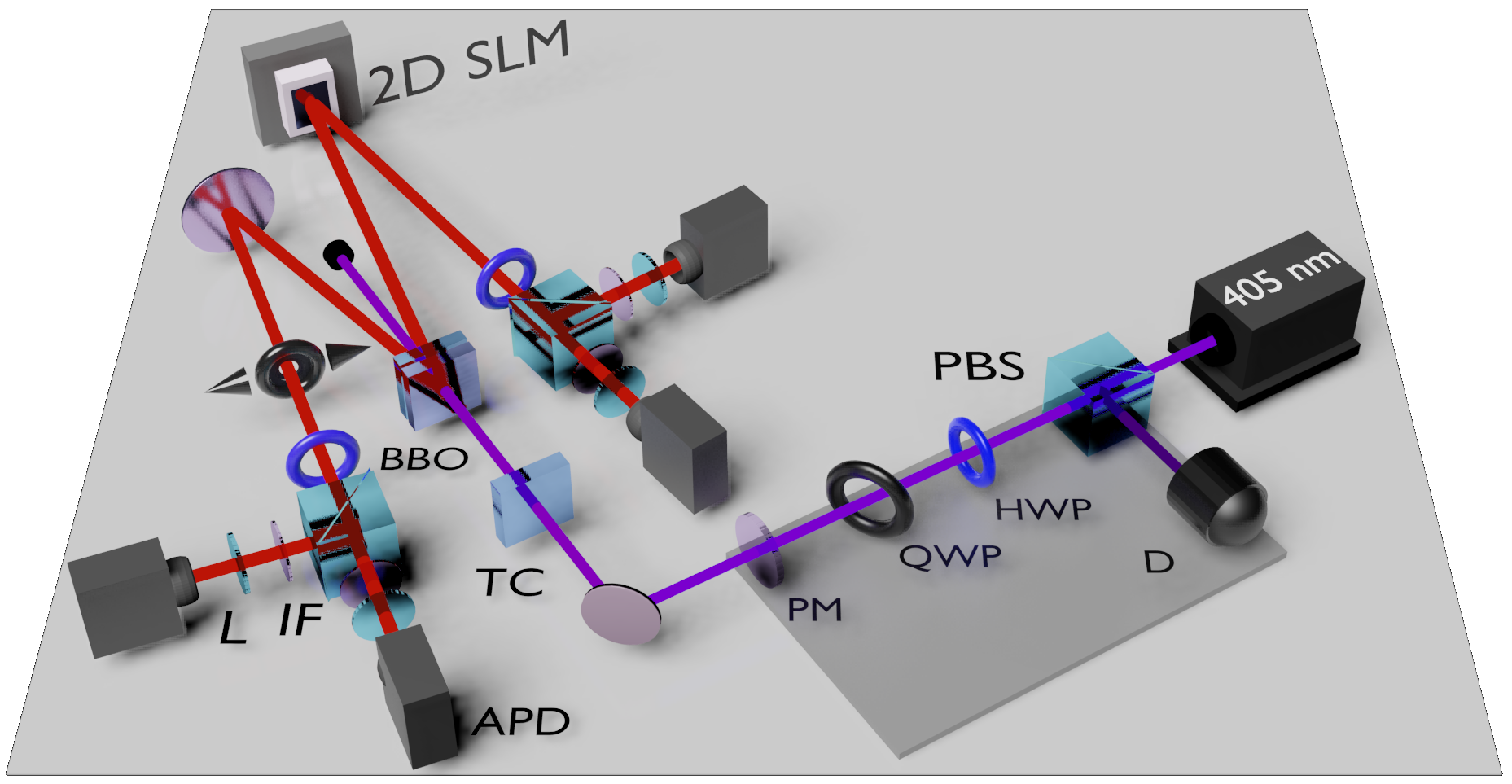}
   \end{tabular}
   \end{center}
   \caption[example] 
   { \label{fig:setup} 
Extended entangled-photons source. The photon pairs are created in two crossed type-I 0.5-mm BBO crystals cut at 29.3$^\mathsf{o}$ and pumped by 405-nm beam to produce noncollinear SPDC photons centered at ~3$^\mathsf{o}$ half-opening angle. A half-wave plate (HWP$_p$) and a tiltable quarter-wave plate (QWP$_p$) manipulate the pump polarization state and are also components of polarization interferometer used for relative-phase measurements.
The relative-phase function is manipulated in frequency using a temporal compensator (TC), and in direction via a two-dimenstional spatial light modulator (2D SLM) charged by the inverted grayscale relative-phase map. PM: Partial mirror; D: Classical detector; APD: Avalanche photodiode, IF: interference filter, L: lens.  
}
\end{figure*}

The coincidence map, on the other hand, is determined by the two-photon wavefunction at the detection plane \cite{klyshko1988photons}
\begin{equation}
\begin{split}
\Psi 
=\langle 0 | E_1^{(+)}(\theta_{x1},\theta_{y1};t_1) E_2^{(+)}(\theta_{x2},\theta_{y2};t_2) |\psi \rangle, \\
\end{split}
\end{equation}
where $|0\rangle$ is the vacuum state and  
\begin{equation*}
\begin{split}
& E_{1}^{(+)}
=\int d\omega~e^{-i\omega t_{1}} \sum_{\mathbf{\sigma}}   
(\mathbf{e}_1 . \mathbf{\sigma}) G(\omega)~ \mathbf{a}_\mathbf{\sigma}(\omega;\theta_{x1},\theta_{y1}), \\
& E_{2}^{(+)}
=\int d\omega~ e^{-i\omega t_{2}}  \sum_{\mathbf{\sigma}}  
(\mathbf{e}_2 . \mathbf{\sigma}) G(\omega)~ \mathbf{a}_\mathbf{\sigma}(\omega;\theta_{x2},\theta_{y2}),
\end{split}
\end{equation*}
are the positive-frequency parts of signal and idler field operators at the space-time coordinates of biphoton detection; $(\theta_{x1},\theta_{y1};t_{1})$ and $(\theta_{x2},\theta_{y2};t_{2})$ \cite{saleh2000duality}. Here the unit vector $\mathbf{e}_i=(e_{ix},e_{iy})$ specifies the orientation of the polarization analyzer in the $i$th SPDC arm, $G(.)$ is the transmissivity profile of the spectral filters, and $\mathbf{a}_\sigma(\omega; \theta_{x},\theta_{y})$ is the photon annihilation operator for a mode of frequency $\omega$, polarization $\mathbf{\sigma}$, and direction defined by the emission angles $(\theta_{x},\theta_{y})$ in free space. The spatiotemporal two-photon wavefunction thus writes \cite{atature2002multiparameter}
\begin{equation}
\begin{split}
\Psi 
= \frac{1}{2} \mathbf{\int}&d\omega _{1}d\omega _{2}
  e^{-i(\omega_1 t_1 + \omega_2 t_2)} 
\\
\times & G(\omega_1) G(\omega_2) \left( e_{1x}e_{2x}\Phi _\mathrm{HH}+ e_{1y}e_{2y} \Phi _\mathrm{VV} \right),
\end{split}
\label{Eq:DetectedWavefn}
\end{equation}
and the probability that the signal photon is captured by the typically used slow-response detector in the direction $(\theta_{x1},\theta_{y1})$ is then given by
\begin{equation}
P(\theta_{x1},\theta_{y1})=\int dt_1 dt_2 d\theta_{x2} d\theta_{y2}|\Psi (\theta_{x1},\theta_{y1},\theta_{x2},\theta_{y2}; t_1, t_2)|^2.
\label{Eq:angdistribution}
\end{equation}
 The schematic of the experimental setup is shown in Fig. \ref{fig:setup}. A collimated 405-nm beam emitted by a continuous-wave diode laser with $\sim$300-fs coherence time is used to pump two abutted 0.5-mm type-I BBO crystals with the optic axis of the first (second) crystal lying in the vertical (horizontal) plane. The two crystals are anti-reflection-coated at all wavelengths of concern and cut at 29.3$^\mathsf{o}$ to produce degenerate SPDC cone centered at ~3$^\mathsf{o}$ half-opening angle.
 The polarizing beam splitter (PBS) improves the extinction ratio of the pump beam (from 20:1 to about 10000:1). The half-wave plate (HWP$_p$) and tiltable quarter-wave plate (QWP$_p$) manipulate the pump polarization state –by rotating the linear polarization and adding phase difference $\phi_p$, respectively– and are also used to conduct the phase measurements as follows. 
 To determine the phase difference $\phi_p$, the partial mirror (PM) back-reflects a tiny portion ($\sim 2\%$) of the pump beam to be analyzed by the dash-outlined polarization interferometer. 
 At its output, the normalized optical power measured by the classical detector (D) is proportionate to $\sin^2 \phi_p$ (see supplementary material). We used 10-nm interference filters centered at 810 nm. This narrowband detection at the degenerate frequency along with the collimated-beam pumping render the approximations: $\theta_{x1}\approx - \theta_{x2}$ and $\theta_{y1}\approx - \theta_{y2}$ reasonable. We will henceforth refer to $\theta_{x1}$ and $\theta_{y1}$ as  $\theta_{x}$ and $\theta_{y}$, respectively, for short.
\begin{figure*} [t]
   \hspace{- .5cm}
   \centering
   \includegraphics[width=17cm]{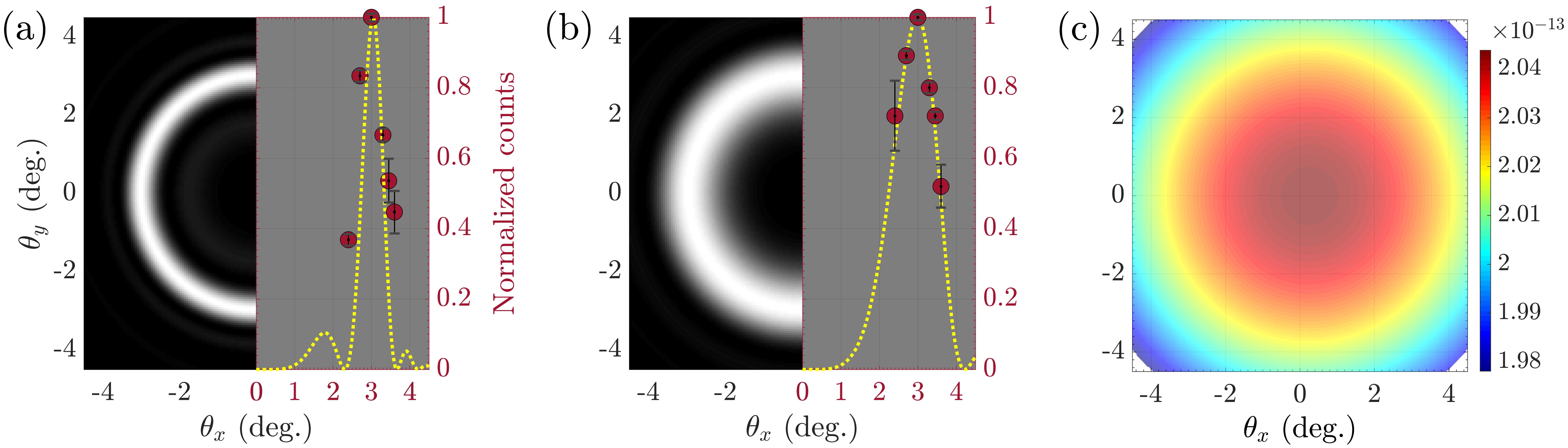}
   \caption[example] 
   { \label{fig:coincidence} 
Coincidence and time-delay maps all over the SPDC cone.
(a) Theoretical prediction (left) and experimental measurements (right) for the angular distribution of biphotons coincidence counts after polarization projection into the diagonal basis state $|++\rangle$. The right inset shows the normalized coincidence rates (red points), experimentally measured behind a 2-mm iris positioned, 50 cm from the SPDC crystals, at the directions: $\theta_x=\{2.4^\mathsf{o}, 2.7^\mathsf{o}, 3^\mathsf{o}, 3.3^\mathsf{o}, 3.45^\mathsf{o}, 3.6^\mathsf{o}\};$ with $\theta_y=0$.  The yellow line shows the normalized rates as predicted by Eq. (\ref{Eq:angdistribution}). (b) The same as in (a) after polarization projection into the rectilinear basis state $|HH\rangle$. 
(c) The directional time-delay map between the HH and VV possibilities for the degenerate biphoton emission as predicted by the derivative of the relative phase $\vartheta$ with respect to $\omega_1$.}
   \end{figure*} 

In the experiment, to figure out the spatial distribution of the scattered biphotons, the coincidence counts are measured behind a 2-mm circular iris translated laterally at $\sim$50 cm from the SPDC crystals. The measurements are made at a number of transverse positions which represent one radial group.
 Figure \ref{fig:coincidence}(a) shows the coincidence counts recorded after polarization projection into the basis state $|++\rangle$. We noticed that this spatial distribution varied when polarization projection was performed with different orientation. Figure \ref{fig:coincidence}(b) shows the coincidence counts after polarization projection into the basis state $|HH\rangle$ which exhibits more angular spread than that of the diagonal projection. This observation is in good agreement with Eq. (\ref{Eq:DetectedWavefn}) and Eq. (\ref{Eq:angdistribution}), where for the projector $|HH\rangle\langle HH|$, $e_{1x}e_{2x}=1$ and $e_{1y}e_{2y}=0$, which yields angular distribution $P(\theta_{x},\theta_{y})\propto \mathrm{sinc}^2 \left(\Delta \kappa^{ooe} L/2\right)$. Alternatively, for the diagonal projector $|++\rangle\langle ++|$, the factors $e_{1x}e_{2x}=e_{1y}e_{2y}=1/2$, therefore $P(\theta_{x},\theta_{y})\propto \mathrm{sinc}^2 \left(\Delta \kappa^{ooe} L/2\right) \cos^2 (\vartheta/2)$. The more confined angular distribution in case of diagonal projection originates from the interference of SPDC biphotons created along the two crystals. This observation shows one side of the coupling effects between polarization and spatial DoFs which can be eliminated via a suitable spatial phase compensation.

\begin{figure} [t]
   \begin{center}
   \begin{tabular}{c} 
   \includegraphics[width=8.5cm]{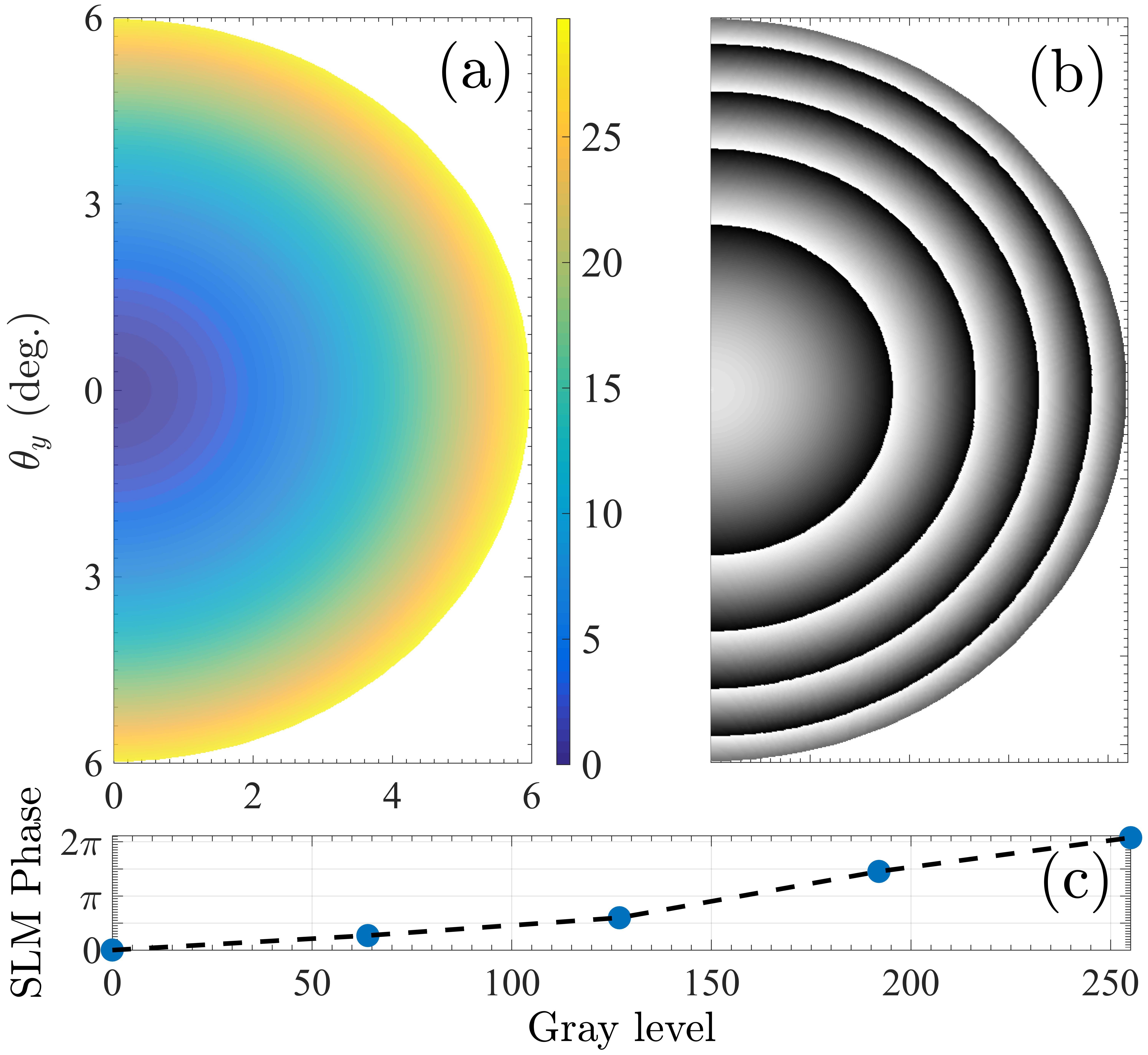}
   \end{tabular}
   \end{center}
   \caption[example] 
   { \label{fig:phasemap} 
(a) The two-dimensional quadratic relative-phase map (in radians) as predicted by Eq. (\ref{eq:relative_phase_approx}) all over the SPDC cone.  (b) Purifying SLM pattern to produce the states $|\phi^\pm\rangle$ with high fidelity over wide emission angles. The pattern constitutes the inverted \textit{modulo-2$\pi$} grayscale of the map in (a). The white rectangle circumscribes the 2D SLM's active area on which the compensation image is loaded. This area covers azimuthal-angle range $\sim$57$^\mathsf{o}$ (of the 180$^\mathsf{o}$ representing half SPDC cone); therefore about one  third  of  the  noncollinear SPDC emission cone could be manipulated.  (c) Experimentally measured phase retardation introduced by the 8-bit SLM at 810 nm plotted versus the gray display level (black:0 and white:255). The phase measurement error ($\pm 0.005 \pi$) is smaller than the readout markers. 
}
   \end{figure}

Figure \ref{fig:coincidence}(c) and Figure \ref{fig:phasemap}(a) show the time-delay and relative-phase maps all over the SPDC scattering cone as obtained by Eq. (\ref{eq:relative_phase_approx}) for the parameters of the experiment.
It can be noticed that while the time-delay map is essentially flat (centered at $\sim$200 fs), the relative-phase map has strong quadratic dependence in the radial direction. 
This time delay is about the same as the coherence time of the SPDC light (determined by the interference filters). Therefore, over that wide directional extent, the temporal distinguishability between the leading (HH) and the lagging (VV) possibilities can be effectively washed out following the conventional approach; adding an offset temporal delay between the parent pump components $|V_p\rangle$ and  $|H_p\rangle$ \cite{nambu2002generation,rangarajan2009optimizing,hegazy2015tunable}. 
In the experiment, this is accomplished using a temporal compensator (TC): Two \textit{available} 0.8-mm 30$^\mathsf{o}$-cut BBO crystals set abutted to each other with their optic axes rotated up-side-down  with respect to each other in the vertical plane; thereby canceling the transverse walk-off. To verify the degree of temporal compensation, we measured the visibility of coincidence fringes after 2-mm iris while HWP$_1$ is rotated and HWP$_2$ remains fixed at $22.5^\mathsf{o}$.
High polarization visibility was observed at all the transverse positions in Fig. \ref{fig:coincidence}; implying the flatness of the time-delay map and the effectiveness of the unified time compensation. 
 At each observation point, the high polarization visibility was achievable at a tilt angle of QWP$_p$ different from its neighboring points (thus, different $\phi_p$). This indicates that the phase-matching part of the relative phase in Eq. (\ref{eq:relative_phase}) varies from a point to another.

Based on this observation, we present here an experimental technique to directly measure the relative phase of the entangled state: 
First, polarization measurements are prepared projecting the two-photon state to diagonal basis states $|++\rangle, $ $|--\rangle$. We then scan the tilt angle of QWP$_p$. Along the scan, reaching the maximum coincidence counts indicates constructing a state nearest to the Bell state: $|\phi^+\rangle= (|H_1H_2\rangle+|V_1V_2\rangle)/\sqrt{2}$ [Here, the zero relative phase indicates that the phase-matching part of the relative phase is complementary to $\phi_p$ as dictated by Eq. (\ref{eq:relative_phase})] \cite{note:auto}. This determines the phase-matching part of the relative phase, since $\phi_p$ value is obtained from the normalized power detected by the photodetector D as aforementioned (see supplementary material). 

\begin{figure} [t]
   \begin{center}
   \begin{tabular}{c} 
   \includegraphics[width=8.5 cm]{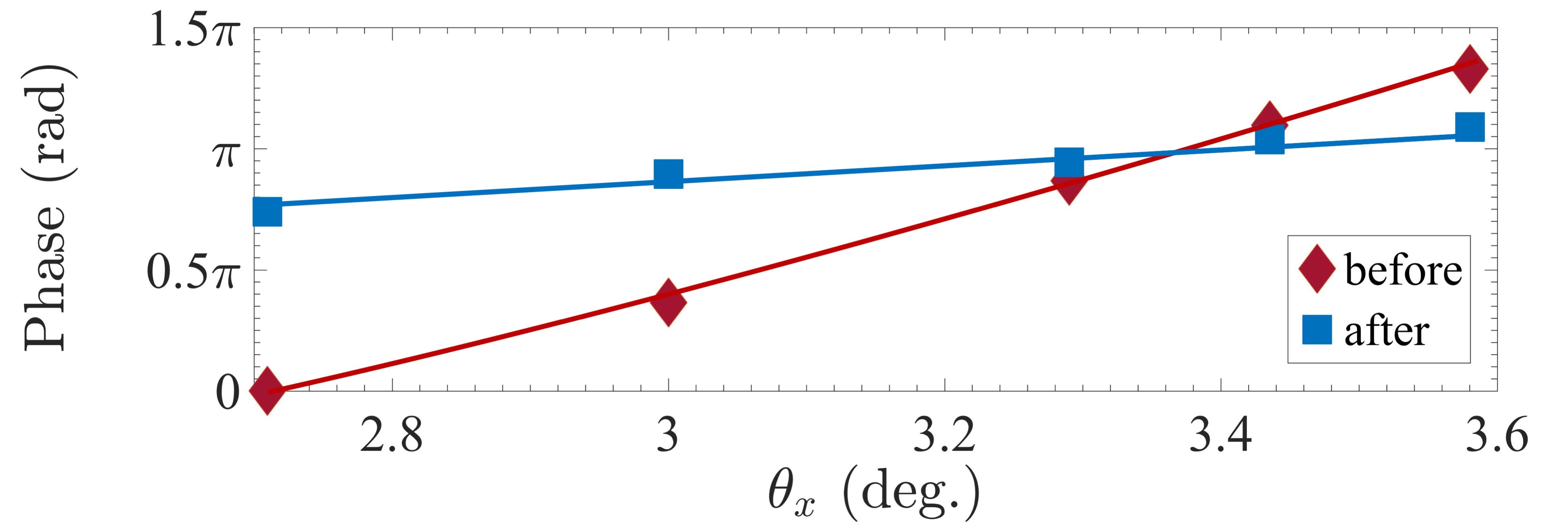}
   \end{tabular}
   \end{center}
   \caption[example] 
   { \label{fig:measuredPhase_gradient} 
Experimentally measured relative phase for the produced polarization-entangled state across the SPDC cone before (brown rhombuses) and after (blue squares) loading the SLM. The measurements are done behind a 2-mm iris positioned, 50 cm from the SPDC crystals, at the directions: $\theta_x=\{2.7^\mathsf{o}, 3^\mathsf{o}, 3.3^\mathsf{o}, 3.45^\mathsf{o}, 3.6^\mathsf{o}\};$ with $\theta_y=0$. The error ($\pm 0.005 \pi$) is smaller than the symbols. Curves are fits to the data. 
}
   \end{figure}

To compensate for the relative-phase variation, a phase-only two-dimensional spatial light modulator (2D SLM) in one of the SPDC arms is used to introduce a pixel-based phase retardation between the H and V polarization components. 
This reflective 2D SLM (display area 15.36 mm$\times$8.64 mm and resolution 1920$\times$1080 pixels) is placed at the transverse plane, $\sim$24.4  cm  from  the  SPDC  crystals, and programmed by a personal computer (PC). After loading by a grayscale image, the SLM develops a phase retardation for each pixel proportional to the pixel's gray level as commonly used in the monochromatic digital display. 
The SLM can introduce up to $\sim 2\pi$ phase shift at 810 nm with its  8-bit display offering 256 intermediate gray/phase levels (black:0 and white:255).
 To treat the relative-phase map, the SLM was charged by the image within the white rectangle in Fig. \ref{fig:phasemap}(b) which is the inverted grayscale image of the modulo-2$\pi$ relative-phase map in Fig. \ref{fig:phasemap}(a). Figure \ref{fig:phasemap}(c) shows the experimentally measured phase of the SLM for the display gray levels \{0 (black), 63, 127, 192, 255 (white)\}.
 A special gamma correction is applied to enhance the linearity of the phase retardation to the pixel gray level at 810 nm.
Figure \ref{fig:measuredPhase_gradient} depicts the experimentally measured relative phase at the transverse positions in Fig. \ref{fig:coincidence} before and after charging the SLM with the compensating image.
The improved flatness of the directional relative-phase function implies less coupling between polarization and spatial DoFs. This yields polarization-entangled state $|\phi^+\rangle$ with high fidelity verified by $\sim 97\%$ polarization visibility measured at all transverse positions in Fig. \ref{fig:coincidence} across the SPDC emission (see supplementary material). This high visibility of coincidence fringes was observed at each  position behind the 2-mm iris while HWP$_1$ was rotated, with HWP$_2$ fixed at $22.5^\mathsf{o}$, and QWP$_p$ set at fixed tilting. By tilting QWP$_p$, the extended high-fidelity emission of the state $|\phi^-\rangle$ could also be formed.

Beside equalizing the quadratic relative phase, the PC-programmed 2D SLM can simultaneously superimpose phase patterns serving other DoFs (e.g., orbital angular momentum modulation/demodulation). However, the following factors have to be taken into account: 
(i) The SLM introduces unavoidable loss factor that can be classified as bulk losses due to the non-ideal reflectivity/transmissivity and diffraction losses due to the pixelation splitting of the incident light into several diffraction orders. In our experiment, the SLM losses reached $\sim 40 \%$. (ii) Effective anti-reflection coating is essential to diminish the possibility that the photon is not manipulated (by reflecting off the first interface) or manipulated multiple times (by rebouncing across the SLM).


To sum up, we have presented an experimental method to compensate for the spatial-spectral relative-phase function of a polarization entangled state, and produce the maximally entangled states $|\phi^\pm\rangle$ with high fidelity over wide emission angle –involving one third of the SPDC cone.
 In frequency domain, the relative-phase gradient has been treated conventionally by the use of unified temporal compensator. 
In momentum domain, the relative phase over extended emission angles exhibits obvious two-dimensional quadratic dependence which has been equalized by using a high resolution PC-programmed 2D SLM loaded by the inverted relative-phase function. We have also developed a simple, yet accurate, experimental approach to measuring the relative phase of the entangled photon pairs by witnessing a normalized portion of the emerging power of the pump beam after polarization interferometry.
The high visibility and flat relative phase of the constructed state has been verified in a set of lateral positions across the SPDC cone. The presented approach can be applied to equalize the entire SPDC emission by using 2D SLM of sufficiently wide active area covering half the SPDC emission cone. Furthermore, the Bell states $|\psi^\pm\rangle$ –extended over wide angles– can be straightforwardly produced by inserting a half-wave plate \cite{note:waveplate} into one SPDC arm.
\\

See supplementary material for further information
about the relative-phase measurement technique and visibility measurements.
\\

This work is supported by ITAC program (Grant No. CFP134), ITIDA, Ministry of Communications and Information Technology (MCIT), Egypt.\\
\\
\noindent\textbf{Data Availability} 

The data that supports the findings of this study are available within the article and its supplementary material.


\begin{thebibliography}{10}
\bibitem{note:auto}
This process can be fully automated by replacing the QWPp by an electrically controlled birefringent element.
\bibitem{note:waveplate}
Such half-wave plate needs to swap the vertical polarization to horizontal
and vice versa over relatively wide angles. It should maintain –at a good precision– the 180$^\mathsf{o}$ phase retardation between its principal axes over wide angles of incidence. The liquid-crystal wave plates –used here for polarization measurements– have generally better performance in this aspect than the traditional quartz wave plates.
\end{thebibliography}
\end{document}